\documentstyle[amssymb,prl,aps,a4]{revtex}

\begin{document}
\draft
\title{Single-electron transistor effect in a two-terminal structure}
\author{S. V. Vyshenski\cite{email}}
\address{Nuclear Physics Institute, Moscow State University,\\
Moscow 119899, Russia; and\\
Research Center for Quantum Effect Electronics,\\
Tokyo Institute of Technology, 2-12-1, Tokyo 152, Japan}
\date{September 2, 1997}
\maketitle

\begin{abstract}
A peculiarity of the single-electron transistor effect makes it possible to
observe this effect even in structures lacking a gate electrode altogether.
The proposed method can be useful for experimental study of charging effects
in structures with an extremely small central island confined between tunnel
barriers (like an ${\simeq}1$ nm quantum dot or a macromolecule probed with
a tunneling microscope), where it is impossible to provide a gate electrode
for control of the tunnel current.
\end{abstract}


By definition, a device called a ``transistor'' should have three terminals.
One of them (the gate) is meant to control the current flowing between the
other two. The same can be said for the case of a single-electron transistor
(SET). The main objective of this paper is to prove that just two terminals
are sufficient for studying the SET effect in experiment, provided that the
voltages applied to these two are held in a special way. Thus in the
particular case of the SET, the transistor effect (TE) can be studied in
systems which are not transistor devices. Although this simplification may
be of no immediate use for the electronics industry, it is of importance for
basic physical experiment. Here interesting and physically rich mesoscopic
systems can be prepared artificially \cite{black} or grown naturally \cite
{sold}. But the nanometer size of these systems makes fabrication of the
gate another challenging problem (if it is feasible at all).

We illustrate the main idea using as an example the semi-classical
``orthodox'' approximation \cite{AvLik} for the description of the SET
dynamics of systems with a purely tunnel conductivity between metallic
electrodes. In the closing section we argue that the same two-terminal
method is much more generally applicable.

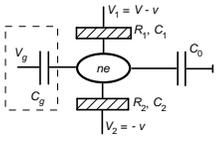
\begin{figure}[tbp]
\caption{Charge-quantized double-barrier structure. Junctions with tunnel
resistances $R_{1,2}$ and capacitances $C_{1,2}$ are shown as boxes. The
self-capacitance $C_0$ of the central island is shown as a capacitor
connected to a point with zero potential. The gate with capacitive coupling $%
C_g$ may be absent from the system.}
\label{scema}
\end{figure}

Consider the charge-quantized double-barrier structure in Fig.\ \ref{scema},
which is called a SET. The total charge $ne$ confined on the central island
is a good macroscopically observable quantum number provided that thermal
and quantum fluctuations of charge are small: $e^2/C\gg k_BT$ and $%
R_{1,2}\gg \hbar /e^2\simeq 4.1$ k$\Omega $, where $C=C_0+C_1+C_2+C_g$.
Traditionally a gate with a capacitive coupling $C_g$ is present and allows
for modulation of the current flowing between terminals $V_1$ and $V_2$. The
modulation is due to the change in charge induced on the central island by a
change in the gate voltage $V_g$. This is the conventional TE \cite{AvLik}.

The gate may be absent from a particular structure. In Fig.\ \ref{scema}
this case is indicated by the dashed lines around the gate. Here we can get
the same modulation effect by making use of a ``hidden'' gate, which is the
self-capacitance $C_0$ of the central island. For this we introduce a common
background $-v$ added to both voltages $V_1$ and $V_2$ simultaneously. We
will see that by changing the voltage $v$ it is possible to observe the same
TE, and for structures on the nanometer scale the efficiency of this $v$
control is approximately the same as would be expected for the best possible
conventional gate.

Thus we are going to exploit an unusual feature of the SET. When it has a
gate (and looks like a 3-terminal structure) it in fact has 4 terminals. The
effective fourth terminal is an infinitely remote point traditionally viewed
as having zero potential. When a SET does not have a regular gate (and looks
like a 2-terminal structure), it is effectively a 3-terminal device, and it
is still possible to observe the TE, this time with a special voltage setup.

\paragraph{Effective additional gate.}

The total charge $ne$ confined on the central island (see Fig.\ \ref{scema})
determines its electrostatic potential $\varphi (n)$: 
\begin{equation}
en+q_{\text{b}}=C\varphi (n)-C_1V_1-C_2V_2-C_gV_g\,.  \label{def}
\end{equation}

Here $q_{\text{b}}$ is a background charge: $q_{\text{b}}/C$ is the
contribution to the potential $\varphi $ of the central island from charged
contaminants present in the vicinity of the island.

Equation\ (\ref{def}) implicitly uses the ``fourth terminal''. The
infinitely remote point used in a definition of the self-capacitance \cite
{Landau} is assumed to be at zero potential. The natural choice [employed in
Eq.\ (\ref{def})] is to have zero potential on an isolated uncharged body.
This choice fixes the gauge. The zero point of the potential is no longer
arbitrary, and the value of the potential (and not just of the potential
difference) acquires absolute meaning.

In other words, the self-interaction of the central island (measured by the
self-capacitance parameter) is equivalent to interaction with a dedicated
point of fixed potential. The most natural choice for such a point is at
infinity (and the natural choice for the fixed potential value is zero). So
the existence of this self-interaction is equivalent to the fact that our
system has a very special point with fixed potential. This special point can
be regarded as a ``hidden'' voltage terminal in our system. We will see that
the voltage parameter $-v$ applied to both current terminals is measured
relative to precisely this hidden terminal. This can be alternatively
regarded as applying a voltage $+v$ to the hidden terminal, which will
imitate one additional $v$ voltage-driven gate.

\paragraph{Orthodox approximation.}

The free-energy costs of increasing ($+$) or decreasing ($-$) the initial
number $n$ of electrons on the central island due to a single-electron
tunneling event ($n\rightarrow n\pm 1$) in junction 1 or 2 are: 
\begin{eqnarray}
F_{1,2}^{\pm }(n) &=&F_f-F_i=\pm e\left[ \varphi (n\pm 1/2)\right] \mp
eV_{1,2}  \nonumber \\
&=&\pm (e/C)(q_{\text{b}}\pm e/2+en+C_gV_g  \nonumber \\
&&+C_1V_1+C_2V_2-CV_{1,2}).  \label{W12}
\end{eqnarray}
where $F_{1,2}^{\pm }<0$ ($>0$) corresponds to an energetically favorable
(unfavorable) event. The dissipation of this energy is part of the tunneling
event and distinguishes macroscopic tunneling (considered here) from
textbook quantum mechanical tunneling. For a given $n$, the tunneling rates
in each junction are expressed \cite{AvLik} by 
\begin{equation}
\Gamma _{1,2}^{\pm }(n)=\frac{-F_{1,2}^{\pm }}{e^2R_{1,2}}\frac 1{1- \exp
\left[ F_{1,2}^{\pm }/(k_BT)\right] }.  \label{rate}
\end{equation}

A statistical distribution $p(n)$ of charge states $n$ is established when
the external voltages are constant. The current $I_i$ through tunnel $i$ in
the direction from $V_1$ to $V_2$ equals 
\begin{equation}
I_{1,2}=\pm e\sum\limits_np\left( n\right) \left[ \Gamma _{1,2}^{+}\left(
n\right) -\Gamma _{1,2}^{-}\left( n\right) \right] ,  \label{current}
\end{equation}
where sum goes over all $n$ for which $p\left( n\right) >0$. Kirchhoff's
law, $I_1=I_2$, holds in the steady state and demands that the distribution $%
p(n)$ should not change in time. More precisely, simultaneous
detailed-balance equations \cite{IngNaz} should hold for all $n$: 
\begin{equation}
p\left( n\right) \Gamma ^{+}\left( n\right) =p\left( n+1\right) \Gamma
^{-}\left( n+1\right) ,  \label{balance}
\end{equation}
with $\Gamma ^{\pm }\left( n\right) =\Gamma _1^{\pm }\left( n\right) +\Gamma
_2^{\pm }\left( n\right) $. For any fixed combination of parameters $C_0$, $%
C_g$, $C_1$, $C_2$, $R_1$, $R_2$, $V_1$, $V_2$, $V_g$, $q_{\text{b}}$, and $T
$, using Eqs.\ (\ref{W12}) and (\ref{rate}), we can solve Eqs.\ (\ref
{balance}) for the statistical distribution $p(n)$. We can then calculate
the current $I$ from Eq.\ (\ref{current}) as a function of these parameters.

\paragraph{Periodic modulation of the current.}

Consider the one-to-one mapping \{$V_1$, $V_2$\} $\leftrightarrow $ \{$v$, $V
$\}: 
\begin{equation}
V_1=V-v,\quad V_2=-v,  \label{bog}
\end{equation}
so that $V_1-V_2=V$ always. In experiment this means that the voltages $V_1$
and $V_2$ are generated [according to Eq.\ (\ref{bog})] by an operational
amplifier or computer starting from two independently controlled parameters: 
$V$ and $v $. By changing $v$ independently of $V$ and other parameters of
the system, we hope to reproduce the TE when the gate is absent completely ($%
C_g=0$).

After applying transformation (\ref{bog}) to Eq.\ (\ref{W12}), we get: 
\begin{equation}
F_{1,2}^{\pm }(n)=\pm (e/C)\left( \pm e/2+K_{1,2}V+en+q\right) ,
\label{W12a}
\end{equation}
with $K_1=-(C_0+C_g+C_2)$, $K_2=C_1$, and partial polarization 
\begin{equation}
q=q_{\text{b}}+C_gV_g+(C_0+C_g)v.  \label{q}
\end{equation}
Recall that the four expressions $F_{1,2}^{\pm }(n)$ determine the
probabilities $p(n)$, current $I$, and all other measurable values.

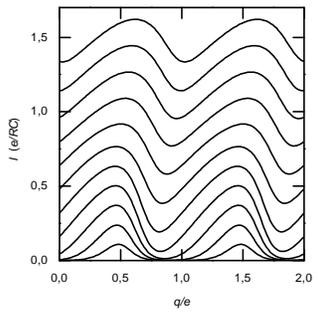
\begin{figure}[tbp]
\caption{Single-electron transistor effect. Current $I$, defined by Eq.\ (%
\ref{current}), versus the effective polarization $q$, defined by Eq.\ (\ref
{q}), at different transport voltages $VC/e$, starting at 0.2 at the bottom
with increments of 0.2. $k_BT=0.05\,e/C$, $C_1=0.7\,C$, $C_2=0.1\,C$, $%
R_2/R_1=10$, $R=R_1+R_2$.}
\label{SETeffect}
\end{figure}

An essential feature of Eq.\ (\ref{W12a}) is that both $q$ and $n$ enter all
four forms $F_{1,2}^{\pm }(n)$ in exactly the same combination $en+q$. As
long as all other parameters of the system are kept constant, the
simultaneous substitutions 
\begin{equation}
\left\{ q\rightarrow q+e,\quad n\rightarrow n-1\right\}  \label{shift}
\end{equation}
leave the combination $en+q$ invariant. So the whole set of [$F_{1,2}^{\pm
}(n)$, $\Gamma _{1,2}^{\pm }(n)$, and $p(n)$] for all $n$ is covariant with
the shift (\ref{shift}). From Eq.\ (\ref{current}) we see that the current $I
$ remians invariant under the change (\ref{shift}). And this just means that
the current is periodic [Fig.\ (\ref{SETeffect})] in $q$ with a period 
\begin{equation}
q_{\text{period}}=e.  \label{qper}
\end{equation}

Note that $K_1$ and $K_2$ in Eq.\ (\ref{W12a}) are always different. They
even have different sign. Therefore, there can be no periodicity in $V$.

In traditional (3-terminal) experiments a monotonic change of $q$ is
achieved through a change of the gate voltage $V_g$. The resulting current
modulation with a period 
\begin{equation}
V_{g\;\text{period}}=e/C_g  \label{gper}
\end{equation}
is known as the single-electron TE.

Alternatively, the same effect can be obtained if the parameter $v$ is
changed with all the other parameters held constant. From Eqs.\ (\ref{q})
and (\ref{qper}) we see that in this case current is modulated with a period 
\begin{equation}
v_{\text{period}}=e/(C_0+C_g).  \label{aper}
\end{equation}

If both parameters $V_g$ and $v$ are changed simultaneously, the current is
modulated with the period (\ref{qper}).

\paragraph{Two-terminal device.}

From Eqs.\ (\ref{W12a}) and (\ref{q}) it is clear that pairs \{$C_g$, $V_g$%
\} and \{$C_0+C_g$, $v$\} play similar roles in SET dynamics. This means
that if the system under study lacks a gate $C_g$ completely ($C_g=0$), one
can still study the TE experimentally, but now with the parameter $C_0$ as
the effective gate, the parameter $v$ as the effective gate voltage, and the
modulation period $v_{\text{period}}$ given by Eq.\ (\ref{aper}).

It can often happen that an interesting two-terminal double-barrier
structure \cite{black} is fabricated in a way which precludes placing a
nearby gate with the sufficiently large $C_g$. Indeed, in demonstrating
periodic modulation of the tunnel current one usually needs to restrict the
voltages to the range $V_g\lesssim 1$ V, just to preserve the mechanical and
electrical stability of the systems under study. Larger voltages may cause
redistribution of the surrounding charged contaminants (changing the
background charge $q_{\text{b}}$) and trigger processes such as
electromigration. To have $V_{g\;\text{period}}\lesssim 1$ V, we need $%
C_g\gtrsim 0.1$ aF. This is hard to achieve for a central island of small
dimensions. If a central island has a radius $r\simeq 1$ nm, as in \cite
{black,sold}, and a gate is separated from it by a distance $d$, then the
gate capacitance can be estimated as $C_g\simeq \varepsilon _{\text{eff}%
}\varepsilon _0\pi r^2/d$. To get $C_g\gtrsim 0.1$ aF, the separation should
be $d\lesssim 2$ nm (with $\varepsilon _{\text{eff}}=10$). It is very hard
to make or find that narrow a separation which is not short-circuited and is
not a tunnel junction. Recall that the typical thickness of a tunnel barrier
is about 1 nm.

This challenging goal was achieved in \cite{sold} by a complicated and
unpredictable method of gate fabrication. The authors began with
lithographic deposition of a gold gate having a highly branched form. The
gate was isolated from the conducting substrate. Then they covered the
structure with a Langmuir film, containing conducting cluster molecules with
radius $r\simeq 1 $ nm. Some (very few) of the clusters happened to lie on
the substrate within a distance $d\lesssim 2$ nm from the gate. Such
clusters were sought out with a scanning tunneling microscope and were then
used as the central island of a SET (substrate---cluster---microscope tip).
This SET was successfully modulated by the gate at room temperature. An
estimate according to Eq.\ (\ref{gper}) gave $C_g=0.2$ aF.

The self-capacitance of a central island with radius $r=1$ nm can be
estimated as $C_0\simeq \varepsilon _{\text{eff}}\varepsilon _0r\simeq 0.1$
aF. And Eq.\ (\ref{aper}) gives $v_{\text{period}}\simeq 1$ V. In real
systems the current leads can screen off some of the environment from the
central island, thus reducing $C_0$ and increasing $v_{\text{period}}$.
However, estimates made for known practical setups always gave a reduction
of $C_0$ by a factor of less than $10$. Thus from Eq.\ (\ref{aper}) we can
expect a value $v_{\text{period}}\simeq 0.3$ V for the same structure. This
means that the authors of \cite{sold} might have demonstrated $v$ modulation
with a period (\ref{aper}) at the same room temperature, even without
fabricating a complicated gate or searching for a cluster molecule which had
accidentally stuck at an appropriate position.

\paragraph{Discussion.}

Consider a SET with a quantum dot as the central island \cite{black}. Due to
spatial quantization of the wave function of an electron confined on the
central island, capacitance parameters $C$ and  $C_0$ are no longer
constants but depend on the charge $ne$, voltages, temperature, and the bulk
and surface properties of the environment \cite{tsukuba}. But even with
variable $C$ and $C_0$, the energy cost of tunneling depends on the
polarization of the central island, and this polarization can be achieved by
changing the voltage $v$ in a two-terminal device. Thus charge quantization
in a quantum-dot SET can be controlled by this effective gate.

Other mechanisms of electron transport (like co-tunneling \cite{AvNaz}, or
thermal activation above the trapping barrier \cite{vyshterm}) may
contribute to the current. In either case the current is periodically
modulated with respect to the polarization of the central island, which in
turn can be achieved by changing either $V_g$ or $v$.

A similar method can be used to control current through charge-quantized
chains of tunnel junctions, in particular, through self-selecting chains of
granules in disordered systems \cite{vyshchan}.


Helpful discussions with F. Ahlers, Y. Nakamura, S. Oda, E. Soldatov, and A.
Zorin are gratefully acknowledged. This work was supported in part by the
Russian Foundation for Basic Research and the Russian Program for Future
Nanoelectronic Devices.


\end{document}